\documentclass[12pt,twoside,fleqn]{article}

\usepackage{a4wide,cite}
\usepackage{amsmath,amssymb}
\usepackage{epsfig,rotating}
\usepackage{axodraw}

\newcommand{\as}{\alpha_s}
\newcommand{\mupi}{\mu_\pi^2}
\newcommand{\mug}{\mu_G^2}
\newcommand{\rd}{\rho_D^3}
\newcommand{\rls}{\rho_{LS}^3}

\newcommand{\bsgamma}{$\bar B\to X_s\gamma$\ }
\newcommand{\MS}{\ensuremath{\overline{\text{MS}}}}

\long\def\symbolfootnote[#1]#2{\begingroup%
\def\thefootnote{\fnsymbol{footnote}}\footnote[#1]{#2}\endgroup}

\numberwithin{equation}{section}
\allowdisplaybreaks[4]

\begin{document}

\begin{titlepage}

\vspace*{3.5cm}

\centerline{\Large\bf\boldmath Power suppressed effects in $\bar B\to X_s\gamma$ at $O(\alpha_s)$}

\vskip 1.5cm

\begin{center}
  {\bf Thorsten Ewerth\symbolfootnote[1]{Present address: Institut f{\"u}r Theoretische Teilchenphysik,
      Karlsruhe Institute of Technology (KIT), D-76128 Karlsruhe, Germany}, Paolo Gambino and Soumitra
    Nandi}\\[2mm]
  {\sl Dip.\ Fisica Teorica, Univ.\ di Torino \& INFN Torino, I-10125 Torino, Italy}
\end{center}

\vskip 2cm

\begin{abstract}
  We compute the $O(\alpha_s)$ corrections to the Wilson coefficients of the dimension five operators emerging
  from the Operator Product Expansion of inclusive radiative $B$ decays. We discuss the impact of the resulting
  $O(\alpha_s \Lambda_{QCD}^2/m_b^2)$ corrections on the extraction of $m_b$ and $\mu_\pi^2$ from the moments of
  the photon spectrum.
\end{abstract}

\end{titlepage}

\section{Introduction}

The inclusive radiative decays of the $B$ meson play a central role in the search for new physics.
While the total rate of \bsgamma is  sensitive to new physics in flavor-changing transitions, its photon spectrum
is almost completely determined by Standard Model physics and can be employed to extract information on the
$B$ meson structure and on the $b$ quark mass, see for instance \cite{Antonelli:2009ws,Buchalla:2008jp}.
The latter are useful in a number of $B$ physics applications, like the determination of $|V_{ub}|$ and of
$|V_{cb}|$, which are both important inputs for the determination of the CKM  unitarity triangle and the study of
CP violation in the Standard Model.

After integrating out the heavy degrees of freedom at the electroweak scale and evolving the resulting weak
Hamiltonian to the $b$ quark mass scale, inclusive radiative and semileptonic $B$ decays are well described by
an Operator
Product Expansion (OPE) in inverse powers of the $b$ quark mass. In this way one can factorize the long distance
dynamics into the matrix elements of a few local operators \cite{Bigi:1992su,Blok:1993va}. Since the Wilson
coefficients of these operators are perturbative, and the matrix elements of the local operators parameterize the
non-perturbative physics, the radiative and semileptonic rates and the moments of their distributions are double
series in $\alpha_s$ and $\Lambda/m_b$, with $\Lambda$ being the QCD scale. The lowest order of this expansion
corresponds to the decay of a free $b$ quark and linear $O(\Lambda/m_b)$ corrections are absent. The relevant
parameters are therefore the heavy quark mass $m_b$ (and possibly $m_c$ for the charmed decays), $\alpha_s$, and
the matrix elements of the local operators: $\mupi$ and $\mug$ at  $O(1/m_b^2)$, $\rd$ and $\rls$ at $O(1/m_b^3)$.
In the case of $b\to s \gamma$ transitions, the first and second central moments of the photon spectrum,
$\langle E_\gamma\rangle$ and $\langle E_\gamma^2-\langle E_\gamma\rangle^2\rangle$, are proportional to the $b$
quark mass and to $\mupi$, respectively, up to subleading corrections in the $1/m_b$ expansion, and contribute
in an important way to the global fits for the extraction of $|V_{cb}|$, $m_b$ and the other OPE parameters
(see \cite{HFAG} for recent results).

The precision of \bsgamma calculations in this framework is known to be limited in two ways: $i)$ The dominant
contribution to the radiative $b$ decay is associated to the electromagnetic dipole operator
$O_7= (\alpha_{\rm em}/4\pi) \,m_b\bar s_L \sigma^{\mu\nu} F_{\mu\nu} b_R$, with  additional operators appearing at
$O(\alpha_s)$. It turns out that a local OPE can be written down for the $O_7$ contribution only, while the other
operators are expected to induce unknown $O(\alpha_s\,\Lambda/m_b)$ contributions, see e.g.\ \cite{paz}; $ii)$ Even for the dominant
$O_7$ component, the experimental cuts employed to isolate the signal introduce a sensitivity to the Fermi motion
of the $b$ quark inside the $B$ meson and tend to disrupt the OPE. One can still resum the higher order
terms into a non-local distribution function \cite{SF} and since the lowest integer moments of this function are
known, one can parameterize it assuming different functional forms \cite{benson}, although alternative approaches
are also possible \cite{Antonelli:2009ws}.

The perturbative corrections to the leading $O_7$ contribution are presently known to complete
Next-to-Next-to-Leading Order (NNLO), i.e.\ $O(\alpha_s^2)$, while the NNLO contributions involving other
operators, quite important for the rate, are not yet complete \cite{Misiak:2006zs}, see \cite{Misiak:2008ss} for
recent updates. As for the non-perturbative corrections to the $O_7$  contribution, they are known through
$O(1/m_b^3)$ \cite{Falk:1993dh,Bauer:1997fe}. Among the non-perturbative corrections associated to operators
other than $O_7$, only one class  is known: they are of $O(1/m_c^2)$ and $O(1/(m_c^2 m_b))$ and numerically
small \cite{Bauer:1997fe}. Overall, the known power corrections modify the total \bsgamma rate by just
$\sim3\%$, but they are essential in the calculation of the moments.

In this paper we present the first calculation of the $O(\alpha_s)$ corrections to the $\Lambda^2/m_b^2$
terms in \bsgamma. The $O(\alpha_s)$ perturbative corrections to the $\mupi$ coefficient in the total rate are
fixed by Lorentz invariance \cite{Bigi:1992su}. Their contribution to the photon moments
has been computed in the context of a multiscale OPE  \cite{neubert} which applies however to
the end-point region only. We have computed the relevant Wilson coefficients at $O(\alpha_s)$ by expanding
off-shell amputated Green functions  around the $b$ quark mass shell, and by matching them onto local operators
in Heavy Quark Effective Theory (HQET). Our results allow for an improved analysis of the
radiative moments. In particular, the inclusion of the $O(\alpha_s)$ perturbative corrections to the variance of
the spectrum, permits the extraction of $\mupi$ at Next-to-Leading-Order (NLO).

The outline of this paper is as follows. In section 2 we recalculate the Wilson coefficients of the matrix
elements of dimension five operators, $\mupi$ and $\mug$, at leading order in $\as$ and introduce our
notation. Section 3 is devoted to the calculation of these Wilson coefficients at $O(\as)$ . It also
contains our final results and a first  numerical estimate of their importance. In section 4 we summarize and
conclude.

\section{Leading Order Coefficients}

At low energy the $b\to s\gamma$ transition is governed by the effective Lagrangian
\begin{equation}\label{eq::lagrangian}
  {\cal L}_\text{eff} =
  {\cal L}_\text{QCD$\times$QED}(u,d,s,c,b) +
  \frac{4G_F}{\sqrt{2}}V_{tb}V_{ts}^* \,C_7^{\rm eff}(\mu)O_7+\dots\,,
\end{equation}
which includes, apart from the QED and QCD interactions of the light quarks $u,\,d,\,s,\,c$ and $b$, the
renormalization scale dependent effective Wilson coefficient $C_7^{\rm eff}$ and the corresponding electromagnetic
operator $O_7$. The latter mediates the $b\to s\gamma$ transition and is given by
\begin{equation}
  O_7 = \frac{\alpha_{\rm em}}{4\pi}\,\overline m_b(\mu)(\bar s\sigma^{\mu\nu}P_Rb)F_{\mu\nu}\,,
\end{equation}
where $\overline m_b(\mu)$ is the running $b$ quark mass in the $\MS$-scheme,
$\sigma_{\mu\nu}=i\,[\gamma_\mu,\gamma_\nu]/2$ and $P_{R,L}=(1\pm\gamma_5)/2$. The ellipses in equation
(\ref{eq::lagrangian}) denote contributions from operators that are not relevant for our calculation.

Using this effective Lagrangian we can derive that part of the differential decay rate that is induced by the
self-interference of the electromagnetic dipole operator,
\begin{equation}
  d\Gamma_{77}(\bar B\to X_s\gamma) =
  \frac{G_F^2\alpha_{\rm em}\overline m_b^2(\mu)}{16\pi^3m_B}\,|V_{tb}V_{ts}^*|^2|C_7^{\rm eff}(\mu)|^2
  \frac{d^3q}{(2\pi)^32E_\gamma}\,W_{\mu\nu\alpha\beta}P^{\mu\nu\alpha\beta}\,.
\end{equation}
Here, $m_B$ is the mass of the $B$ meson, $q$ the momentum of the photon, $W_{\mu\nu\alpha\beta}$ the
hadronic tensor,
\begin{align}\label{eq::hadronictensor}
  W_{\mu\nu\alpha\beta} &= \sum_X\int\!\frac{d^3p_X}{(2\pi)^32E_X}
  (2\pi)^4\delta^{(4)}(p_B-p_X-q)\nonumber\\[2mm]
  &\qquad\times\,\langle\bar B(p_B)|\bar b(0)\sigma_{\mu\nu}P_Ls(0)|X_s(p_X)\rangle
  \langle X_s(p_X)|\bar s(0)\sigma_{\alpha\beta}P_Rb(0)|\bar B(p_B)\rangle\,,
\end{align}
and $P^{\mu\nu\alpha\beta}$ the photon tensor,
\begin{align}
  P^{\mu\nu\alpha\beta} &= \sum_{\lambda=\pm 1}\langle 0|F^{\mu\nu}|\gamma(q,\lambda)\rangle
  \langle \gamma(q,\lambda)|F^{\alpha\beta}|0\rangle\nonumber\\[1mm]
  &=q^\alpha q^\nu g^{\mu\beta}-q^\alpha q^\mu g^{\nu\beta} +
  q^\mu q^\beta g^{\nu\alpha}-q^\beta q^\nu g^{\mu\alpha}\,.
\end{align}
The hadronic tensor $W_{\mu\nu\alpha\beta}$ itself can be rewritten as
\begin{align}
  W_{\mu\nu\alpha\beta} = 2\,{\rm Im}\left(i\int\!d^4x\,e^{-iq\cdot x}
  \langle\bar B(p_B)|\,T\!\left\{\bar b(x)\sigma_{\mu\nu}P_Ls(x)
  \bar s(0)\sigma_{\alpha\beta}P_Rb(0)\right\}|\bar B(p_B)\rangle\right)\,,
\end{align}
which is useful because the time-ordered product can be expanded into a series of local operators that are
suppressed by powers of the $b$ quark mass. Hence we can write
\begin{equation}\label{eq::ope}
  W_{\mu\nu\alpha\beta}P^{\mu\nu\alpha\beta} =
  -16\pi m_b\Big(c_{\rm dim\,3}O_{\rm dim\,3}+\frac{1}{m_b}\,c_{\rm dim\,4}O_{\rm dim\,4} +
  \frac{1}{m_b^2}\,c_{\rm dim\,5}O_{\rm dim\,5}+\dots\Big)\,,
\end{equation}
where $O_{{\rm dim}\,n}$ is an operator of dimension $n$ that contains $n-3$ derivatives, and
\begin{equation}
  c_{{\rm dim}\,n} = c_{{\rm dim}\,n}^{(0)}+\frac{\alpha_s}{4\pi}\,c_{{\rm dim}\,n}^{(1)}+\dots
\end{equation}
is the corresponding Wilson coefficient that can be determined in perturbation theory. In the reminder of
this section we will review the calculation of $c^{(0)}_{{\rm dim}\,n}$ for $n=3,\,4,\,5$.

\begin{figure}[t]
  \vspace*{2cm}
  \begin{center}
    \begin{tabular}{ccc}
      \begin{picture}(0,0)(0,0)
        \SetScale{.8}
        % diagram
        \SetWidth{1.8}
        \Line(-100,-55)(-40,0)
        \Line(100,-55)(40,0)
        \SetWidth{.5}
        \Line(-40,0)(40,0)
        \Photon(-100,55)(-40,0){2}{5}
        \Photon(100,55)(40,0){2}{5}
        % momenta
        \Text(-45,-32)[br]{$\nearrow$}
        \Text(-18,-35)[br]{\scriptsize $m_bv+k$}
        \Text(-46,24)[br]{$\nwarrow$}
        \Text(-42,28)[br]{\scriptsize $q$}
        \Text(57,22)[br]{$\swarrow$}
        \Text(51,31)[br]{\scriptsize $q$}
        \Text(72,-25)[br]{$\searrow$}
        \Text(100,-18)[br]{\scriptsize $m_bv+k$}
        % quark labels
        \Text(-78,-39)[br]{\small $b$}
        \Text(78,-39)[bl]{\small $b$}
        \Text(0,-8)[bl]{\small $s$}
        % operators
        \Text(-35,0)[c]{\rule{2mm}{2mm}}
        \Text(-43,-5)[br]{\small $O_7$}
        \Text(33,0)[c]{\rule{2mm}{2mm}}
        \Text(43,-5)[bl]{\small $O_7$}
      \end{picture}
      & \hspace{6.8cm} &
      \begin{picture}(0,0)(0,0)
        \SetScale{.8}
        % diagram
        \SetWidth{1.8}
        \Line(-100,-55)(-40,0)
        \Line(100,-55)(40,0)
        \SetWidth{.5}
        \Line(-40,0)(40,0)
        \Photon(-100,55)(-40,0){2}{5}
        \Photon(100,55)(40,0){2}{5}
        \Gluon(0,0)(0,55){3}{5}
        % momenta
        \Text(-45,-32)[br]{$\nearrow$}
        \Text(-18,-35)[br]{\scriptsize $m_bv+k$}
        \Text(-46,24)[br]{$\nwarrow$}
        \Text(-42,28)[br]{\scriptsize $q$}
        \Text(57,22)[br]{$\swarrow$}
        \Text(51,31)[br]{\scriptsize $q$}
        \Text(12,18)[br]{$\uparrow$}
        \Text(17,19)[br]{\scriptsize $r$}
        \Text(72,-25)[br]{$\searrow$}
        \Text(114,-18)[br]{\scriptsize $m_bv+k-r$}
        % quark labels
        \Text(-78,-39)[br]{\small $b$}
        \Text(78,-39)[bl]{\small $b$}
        \Text(0,-8)[bl]{\small $s$}
        % operators
        \Text(-35,0)[c]{\rule{2mm}{2mm}}
        \Text(-43,-5)[br]{\small $O_7$}
        \Text(33,0)[c]{\rule{2mm}{2mm}}
        \Text(43,-5)[bl]{\small $O_7$}
      \end{picture}\\[12mm]
    \end{tabular}
  \end{center}
  \caption{\sl The imaginary parts of these tree-level diagrams contribute to the Wilson coefficients of the
    operators with dimension 3, 4 and 5. }\label{fig::tree-level}
\end{figure}
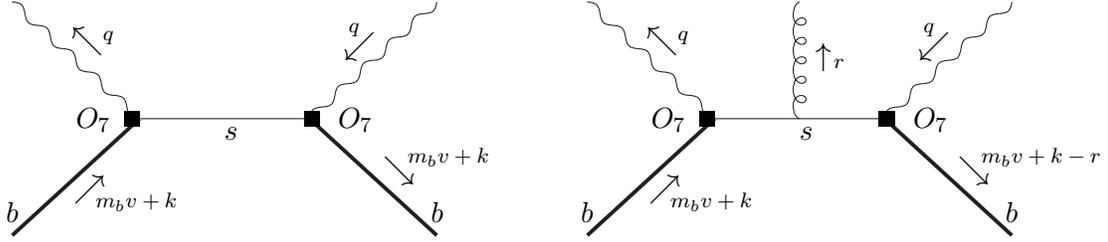

The two Feynman diagrams contributing to these Wilson coefficients are depicted in Fig.~\ref{fig::tree-level}. The
momentum of the incoming $b$ quark is $m_bv+k$, where $v$ is the velocity of the $B$ meson, and $k$ is a residual
momentum that accounts for the interaction of the almost on-shell $b$ quark with the light degrees of freedom in
the $B$ meson. The components of this residual momentum are of  $O(\Lambda)$, i.e.\  much smaller than the
$b$ quark mass. The same is true for the components of the momentum $r$ of the radiated soft gluon that is present
in the  Feynman diagram on the right-hand side. To achieve the anticipated OPE (\ref{eq::ope}) we perform a Taylor
expansion of the amputated Green functions corresponding to the two tree-level diagrams up to $O(k^2)$ for the
left diagram and up to $O(k)$ and $O(r)$ for the right diagram. This is equivalent to an expansion in inverse 
powers of $m_b$, or more precisely, to an expansion in $(iD_\mu-m_b v_\mu)/m_b$, where
$D_\mu=\partial_\mu+ig_sG_\mu^aT^a$. In the algebraic manipulations we refrain from using any on-shell relations for
the $b$ quark, that is we do not impose any restriction on the residual momenta. The only on-shell condition we
apply is that of the $B$ meson, namely $v^2$$\,=\,$$1$. After extracting the imaginary part of the amputated Green
functions we replace the residual momenta $k$ and $r$ with derivatives acting on the $b$ and the gluon fields,
interpreting the latter as parts of a covariant derivative. We end up with expressions for the two tree-level
diagrams in terms of local operators of the form $\bar b F(v,q,D)b$, where $F(v,q,D)$ is a matrix-valued function
of $v_\mu$, $q_\mu$ and $D_\mu$ in spinor space. As far as the $s$-quark is concerned, we will use $m_s=0$.

At this point it is convenient to introduce HQET \cite{Georgi:1990um} which incorporates the expansion in
$(iD_\mu-m_bv_\mu)/m_b$ in a natural manner. Its Lagrangian is given by \cite{Neubert:1993mb,Manohar:2000dt}
\begin{equation}
  {\cal L}_{\rm HQET}= i\bar b_v v\cdot D b_v + \frac1{2m_b} \bar b_v (i D_\perp)^2 b_v -
  a(\mu) \frac{g_s}{4 m_b} \bar b_v \sigma_{\mu\nu} G^{\mu\nu}  b_v + O\left(\frac1{m_b^2}\right)\,,
\end{equation}
where $D_\perp^\mu=D^\mu- v ^\mu \,v\cdot D$ and \cite{Eichten:1990vp}
\begin{equation}
  a(\mu) = 1 + \left[C_F+C_A\left(1+\ln\frac{\mu}{m_b}\right)\right]\frac{\alpha_s}{4\pi}+ \dots\,.
\end{equation}
The numerical values of the color factors are $C_F=4/3$ and $C_A=3$. The relation between the $b$ quark fields in
QCD and in HQET reads
\begin{equation}\label{eq::hqetspinor}
  b(x) = e^{-im_bv\cdot x}\left(1+\frac{iD\!\!\!\!/_\perp}{2m_b}\right)b_v(x) + O\left(\frac{1}{m_b^2}\right)\,,
\end{equation}
at tree-level as well as at $O(\alpha_s)$.

Using (\ref{eq::hqetspinor}), together with $v\!\!\!/b_v=b_v$, in all local operators of dimension 4 and 5, we
end up with the following set of operators,
\begin{align}\label{eq::operators}
  O_b^\mu &= \bar b\gamma^\mu b, &
  O_2^{\mu\nu} &= \bar b_v\frac{1}{2}\{iD^\mu,iD^\nu\}b_v\,, &&&\nonumber\\[1mm]
  O_1^{\mu} &= \bar b_viD^\mu b_v, &
  O_3^{\mu\nu} &= \bar b_v\frac{g_s}{2}G^{a\mu}_{\;\;\;\;\alpha}\sigma^{\alpha\nu}T^ab_v\,, &&&
\end{align}
where $g_sG^a_{\mu\nu}T^a=-i[D_\mu,D_\nu]$. We also find  operators that include a $\gamma_5$. However, we will
eventually calculate  matrix elements of operators between $B$ meson states. From parity considerations it follows
then that only the operators given in (\ref{eq::operators}) give non-vanishing contributions. Hence it is not
necessary to calculate the Wilson coefficients of those operators that include a $\gamma_5$. We should also mention
that up to this point of the calculation we did not use the equation of motion for the $b_v$ field that follows
from the HQET Lagrangian.

For the Wilson coefficients of the operators given in (\ref{eq::operators}) we find, in
$d=4-2\epsilon$ space-time dimensions,
\begin{align}
  c_{b\mu}^{(0)} &= -\frac{1}{2}\,(1-\epsilon)\,\hat q_\mu\delta(1-z)\,,\nonumber\\[1mm]
  c_{1\mu}^{(0)} &= -\frac{1}{2}\,(1-\epsilon)\Big[\left(2 v_\mu-\hat q_\mu\right)\delta(1-z) +
    \left(v_\mu-\hat q_\mu\right)\delta^\prime(1-z)\Big],\nonumber\\[1mm]
  c_{2\mu\nu}^{(0)} &= -\frac{1}{4}\,(1-\epsilon)\Big[
    2\left(g_{\mu\nu}+v_\mu\hat q_\nu\right)\delta(1-z) +
    \left(g_{\mu\nu}+6v_\mu v_\nu-6v_\mu\hat q_\nu\right)\delta^\prime(1-z)\nonumber\\[1mm]
  &\hspace{27mm}+ 2\left(v_\mu v_\nu-2v_\mu\hat q_\nu+\hat q_\mu\hat q_\nu\right)
    \delta^{\prime\prime}(1-z)\Big],\nonumber\\[1mm]
  c_{3\mu\nu}^{(0)} &= -\Big[2v_\mu\hat q_\nu-\frac{1+\epsilon}2 g_{\mu\nu}\Big]\delta(1-z) +
  \Big[(3+\epsilon)\hat q_\mu\hat q_\nu-2v_\mu\hat q_\nu +
    \frac{1+\epsilon}{4}g_{\mu\nu}\Big]\delta^\prime(1-z)\,,
\end{align}
where $z=2\,v\cdot\hat q=2E_\gamma/m_b$ and $\hat q=q/m_b$.

In the last step we have to calculate the forward matrix elements of the four operators given in
(\ref{eq::operators}) between $B$ meson states. Since we expressed the leading operator $O_b^\mu$ in terms
of the $b$ quark fields of QCD we know its matrix element exactly,
\begin{equation}\label{eq::lo}
  \langle\bar B(p_B)|O_b^\mu|\bar B(p_B)\rangle = 2 m_B v^\mu\,.
\end{equation}
The evaluation of the matrix elements of the other operators involves the equation of motion of the effective
theory, and leads to two additional matrix elements \cite{Falk:1992wt},
\begin{equation}\label{eq::lambdai}
  \lambda_1 = \frac{1}{2m_B}\langle\bar B(v)|\bar b_v(iD)^2b_v|\bar B(v)\rangle\,,\qquad
  \lambda_2 = -\frac{1}{6m_B}\langle\bar B(v)|\bar b_v\frac{g_s}{2}G_{\mu\nu}\sigma^{\mu\nu}b_v|\bar B(v)\rangle\,.
\end{equation}
The velocity dependent $B$ meson states used here are related to the momentum dependent
ones introduced in (\ref{eq::hadronictensor}) by
$|\bar B(p_B)\rangle=\sqrt{m_B}\,|\bar B(v)\rangle+O(1/m_b)$. The power correction to this relation is irrelevant
for our calculation.
While $\lambda_1$ and $\lambda_2$ are defined in the asymptotic HQET regime, in practical
applications one deals
with $\mu_\pi^2= -\lambda_1+O(1/m_b)$ and $\mu_G^2=3\lambda_2+O(1/m_b)$, defined  in terms of the $m_b$-finite
QCD states appearing in (\ref{eq::hadronictensor}).

Now we are in a position to calculate the contribution to the differential decay rate due to the electromagnetic
dipole operator only. We obtain
\begin{align}\label{eq::decayrate}
  \frac{d\Gamma_{77}}{dz} &= \Gamma_{77}^{(0)}\Bigg[ 
  c_{0}^{(0)}+c_{\lambda_1}^{(0)}\frac{\lambda_1}{2m_b^2} +
    c_{\lambda_2}^{(0)}\frac{\lambda_2(\mu)}{2m_b^2} +
    \frac{\as(\mu)}{4\pi}\left(c_{0}^{(1)}+c_{\lambda_1}^{(1)}\frac{\lambda_1}{2m_b^2} +
    c_{\lambda_2}^{(1)}\frac{\lambda_2(\mu)}{2m_b^2}\right)\Bigg]\,,
\end{align}
where
\begin{equation}
  \Gamma_{77}^{(0)} = \frac{G_F^2\alpha_{\rm em}\bar m_b^2(\mu)m_b^3}{32\pi^4}\,|V_{tb}V_{ts}^*|^2
  |C_7^{\rm eff}(\mu)|^2\,,
\end{equation}
and
\begin{align}
  c_{0}^{(0)} &= \delta(1-z)\,,\qquad
  c_{\lambda_1}^{(0)} = \delta(1-z)-\delta^\prime(1-z)-\frac{1}{3}\,\delta^{\prime\prime}(1-z)\,,\nonumber\\[1mm]
  c_{\lambda_2}^{(0)} &= -9\,\delta(1-z)-3\,\delta^\prime(1-z)\,.
\end{align}
This is in agreement with the well-known results given in \cite{Falk:1993dh}. The calculation of the Wilson
coefficients $c_{0}^{(1)}$, $c_{\lambda_1}^{(1)}$ and $c_{\lambda_2}^{(1)}$ is the subject of the next section.

\section{Next-to-Leading Order Coefficients}

\begin{figure}[t]
  \vspace*{2cm}
  \begin{center}
    \begin{tabular}{ccc}
      \begin{picture}(0,0)(0,0)
        \SetScale{.8}
        % diagram
        \SetWidth{1.8}
        \Line(-100,-55)(-40,0)
        \Line(100,-55)(40,0)
        \SetWidth{.5}
        \Line(-40,0)(40,0)
        \Photon(-100,55)(-40,0){2}{5}
        \Photon(100,55)(40,0){2}{5}
        \GlueArc(-40,0)(40,-138,0){5}{6}
        % momenta
        \Text(-78,-39)[br]{\small $b$}
        \Text(78,-39)[bl]{\small $b$}
        \Text(12,8)[tl]{\small $s$}
        % operators
        \Text(-35,0)[c]{\rule{2mm}{2mm}}
        \Text(-43,-5)[br]{\small $O_7$}
        \Text(33,0)[c]{\rule{2mm}{2mm}}
        \Text(43,-5)[bl]{\small $O_7$}
      \end{picture}
      & \hspace{6.8cm} &
      \begin{picture}(0,0)(0,0)
        \SetScale{.8}
        % diagram
        \SetWidth{1.8}
        \Line(-100,-55)(-40,0)
        \Line(100,-55)(40,0)
        \SetWidth{.5}
        \Line(-40,0)(40,0)
        \Photon(-100,55)(-40,0){2}{5}
        \Photon(100,55)(40,0){2}{5}
        \GlueArc(40,0)(40,-180,-42){5}{6}
        % momenta
        \Text(-78,-39)[br]{\small $b$}
        \Text(78,-39)[bl]{\small $b$}
        \Text(-12,8)[tr]{\small $s$}
        % operators
        \Text(-35,0)[c]{\rule{2mm}{2mm}}
        \Text(-43,-5)[br]{\small $O_7$}
        \Text(33,0)[c]{\rule{2mm}{2mm}}
        \Text(43,-5)[bl]{\small $O_7$}
      \end{picture}\\[35mm]
      \begin{picture}(0,0)(0,0)
        \SetScale{.8}
        % diagram
        \SetWidth{1.8}
        \Line(-100,-55)(-40,0)
        \Line(100,-55)(40,0)
        \SetWidth{.5}
        \Line(-40,0)(40,0)
        \Photon(-100,55)(-40,0){2}{5}
        \Photon(100,55)(40,0){2}{5}
        \GlueArc(0,50)(110,-132,-48){5}{9}
        % momenta
        \Text(-78,-39)[br]{\small $b$}
        \Text(78,-39)[bl]{\small $b$}
        \Text(0,8)[tl]{\small $s$}
        % operators
        \Text(-35,0)[c]{\rule{2mm}{2mm}}
        \Text(-43,-5)[br]{\small $O_7$}
        \Text(33,0)[c]{\rule{2mm}{2mm}}
        \Text(43,-5)[bl]{\small $O_7$}
      \end{picture}
      & \hspace{6.8cm} &
      \begin{picture}(0,0)(0,0)
        \SetScale{.8}
        % diagram
        \SetWidth{1.8}
        \Line(-100,-55)(-40,0)
        \Line(100,-55)(40,0)
        \SetWidth{.5}
        \Line(-40,0)(40,0)
        \Photon(-100,55)(-40,0){2}{5}
        \Photon(100,55)(40,0){2}{5}
        \GlueArc(0,0)(20,-180,0){5}{4}
        % momenta
        \Text(-78,-39)[br]{\small $b$}
        \Text(78,-39)[bl]{\small $b$}
        \Text(0,8)[tl]{\small $s$}
        % operators
        \Text(-35,0)[c]{\rule{2mm}{2mm}}
        \Text(-43,-5)[br]{\small $O_7$}
        \Text(33,0)[c]{\rule{2mm}{2mm}}
        \Text(43,-5)[bl]{\small $O_7$}
      \end{picture}\\[16mm]
    \end{tabular}
  \end{center}
  \caption{\sl One-loop diagrams contributing to the Wilson coefficients of the
    operators with dimension 3, 4 and 5.
    The momenta assignments of the external lines are the same as in
    figure \ref{fig::tree-level}.
    Sixteen additional diagrams with a gluon radiated 
    off an internal line can be obtained from the diagrams shown here.
}\label{fig::oneloop}
\end{figure}
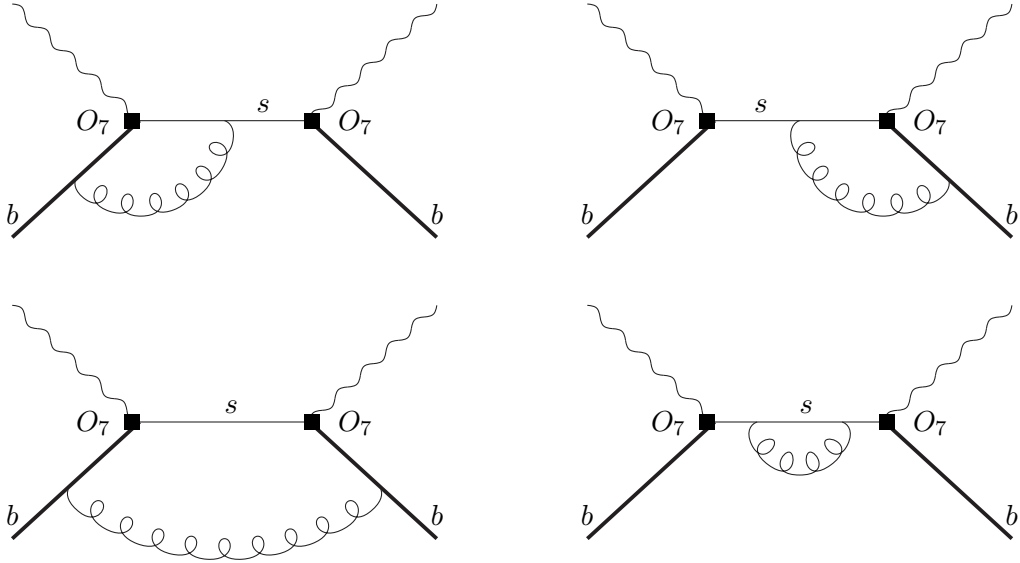

In order to determine the Wilson coefficients $c_{0}^{(1)}$, $c_{\lambda_1}^{(1)}$ and $c_{\lambda_2}^{(1)}$
we calculate the amputated Green functions corresponding to the Feynman diagrams shown in Fig.~\ref{fig::oneloop}.
The momentum assignments of the external lines are exactly the same as in the tree-level calculation performed in
the last section. For the Green functions that contain a radiated soft gluon we apply the background-field
formalism \cite{Abbott:1980hw}. Furthermore, we work in the general $R_\xi$-gauge for the gluon propagator. Again we
perform a Taylor expansion up to $O(k^2)$ for the diagrams without a radiated soft gluon and up to $O(k)$ and $O(r)$
for the diagrams with a radiated soft gluon, and refrain from using any on-shell relations for the $b$ quark.
We apply integration-by-parts techniques \cite{Laporta:2001dd} to reduce all integrals to a few so-called master
integrals and solve the latter analytically. Ultraviolet as well as infrared divergences are handled by
dimensional regularization. The ultraviolet divergences can be removed by appropriate
renormalization. For the self-mixing of the operator $O_7$ we use the $\MS$-scheme,
\begin{equation}
  Z_{m_b}^{\MS}Z_{77}^{\MS} = 1 + \frac{C_F}{\epsilon}\frac{\as}{4\pi} + \dots\,,
\end{equation}
and for the field renormalization constant of the $b$ quark we apply the on-shell scheme,
\begin{equation}
  Z_{b}^{\rm OS} = 1 - C_F\left(\frac{3}{\epsilon}+4+6\ln\frac{\mu}{m_b}\right)\frac{\as}{4\pi} + \dots\,.
\end{equation}
As far as the renormalization of the background gluon field $G_\mu^a$ is concerned, we only have to
remember that $g_sG_\mu^a$ does not get renormalized. Finally, we extract the imaginary part of the amputated
Green functions, replace the residual momenta $k$ and $r$ with covariant derivatives $iD_\mu-m_b v_\mu$ and
the $b$ quark spinors with $b_v$ spinors via (\ref{eq::hqetspinor}), and calculate the forward matrix
elements of all operators between $B$ meson states. Our result is then a linear combination of the three matrix
elements introduced in (\ref{eq::lo}) and (\ref{eq::lambdai}) with coefficients that are ultraviolet finite
but still contain infrared divergences,
\begin{equation}\label{eq::lhs}
  f_0^\mu\left(z,\xi,\mu,\frac{1}{\epsilon_{IR}}\right)v_\mu +
  f_{\lambda_1}\left(z,\xi,\mu,\frac{1}{\epsilon_{IR}}\right)\frac{\lambda_1}{2m_b} +
  f_{\lambda_2}\left(z,\xi,\mu,\frac{1}{\epsilon_{IR}}\right)\frac{\lambda_2}{2m_b}\,.
\end{equation}
We have also made explicit the dependence on the gauge parameter $\xi$. The infrared divergences are removed in
the matching procedure.

We now turn our attention to the right-hand side of the matching equation (\ref{eq::ope}). At the one-loop
level it schematically looks like
\begin{equation}\label{eq::rhs}
  -16\pi m_b\sum_{n=3}^\infty\frac{1}{m_b^{n-3}}\bigg[
  c_{\rm dim\,n}^{(0)}\langle O_{\rm dim\,n}\rangle_{\mbox{\scriptsize 1-loop}} +
  \bigg(\frac{\alpha_s}{4\pi}\,c_{\rm dim\,n}^{(1)}+\delta Z_{\rm dim\,n}^{}c_{\rm dim\,n}^{(0)}\bigg)
  \langle O_{\rm dim\,n}\rangle_{\mbox{\scriptsize tree}}\bigg]\,,
\end{equation}
where $\langle O_{\rm dim\,n}\rangle_{\mbox{\scriptsize tree}}$ and
$\langle O_{\rm dim\,n}\rangle_{\mbox{\scriptsize 1-loop}}$ denote the tree-level and one-loop matrix elements of
the operator $O_{\rm dim\,n}$ between $B$ meson states, respectively, and $Z_{\rm dim\,n}=1+\delta Z_{\rm dim\,n}$
collects the $Z$-factors to render this expression ultraviolet finite. In the case at hand we only have to
consider the one-loop matrix elements of the operators given in (\ref{eq::operators}) since these are the only
ones that have non-vanishing Wilson coefficients at the tree-level. The same holds for the tree-level matrix elements
that are multiplied by $Z$-factors. 
Because of the Taylor expansion in the momenta $k$ and $r$ the 
one-loop matrix elements of the operators $O_2^\mu$ and $O_{3,4}^{\mu\nu}$ vanish in dimensional regularization and
we only need to compute the one-loop matrix elements of $O_b^\mu$. The
operators that include a $\gamma_5$ can again be discarded due
to parity considerations. For the renormalization constants, we use the on-shell scheme for the $b$ and $b_v$
spinors, and the $\MS$ scheme for the operator renormalization,
\begin{align}
  \left[c_{b\mu}O_b^\mu\right]^{\rm bare} &= Z_b^{\rm OS}\,c_{b\mu}O_b^\mu\,, &
  \left[c_{2\mu\nu}O_2^{\mu\nu}\right]^{\rm bare} &=
    Z_{b_v}^{\rm OS}Z_{\rm kin}^{\MS,\mu\nu\alpha\beta}\,c_{2\mu\nu}O_{2\alpha\beta}\,, &&\nonumber\\[1mm]
  \left[c_{1\mu}O_1^\mu\right]^{\rm bare} &= Z_{b_v}^{\rm OS}\,c_{1\mu}O_1^\mu\,, &
  \left[c_{3\mu\nu}O_3^{\mu\nu}\right]^{\rm bare} &=
    Z_{b_v}^{\rm OS}Z_{\rm chromo}^{\MS,\mu\nu\alpha\beta}\,c_{3\mu\nu}O_{3\alpha\beta}\,.
\end{align}
A simple one-loop calculation yields
\begin{align}
  Z_{\rm kin}^{\MS,\mu\nu\alpha\beta} &= -C_F \frac{3-\xi}{\epsilon}
  \left(g^{\mu\nu}-2v^\mu v^\nu\right)v^\alpha v^\beta\,\frac{\as}{4\pi}+\dots\,\nonumber\\[1mm]
  Z_{\rm chromo}^{\MS,\mu\nu\alpha\beta} &= \frac{C_A}{\epsilon}
  \left(g^{\mu\alpha}-v^\mu v^\alpha\right)g^{\nu\beta}\,\frac{\as}{4\pi}+\dots\,.
\end{align}
The Feynman gauge is obtained by setting $\xi=1$. The on-shell renormalization constant $Z_{b_v}^{\rm OS}$ can be
set equal to 1 since the $O(\alpha_s)$-corrections to the selfenergy of the $b_v$ field depends on no other scale
than the renormalization scale $\mu$.

Requiring equality between (\ref{eq::lhs}) and (\ref{eq::rhs}), and solving for $c_{\rm dim\,n}^{(1)}$, we obtain
infrared finite and gauge independent expressions. Writing the coefficients as follows,
\begin{align}\label{eq::results}
  c_{0}^{(1)} &= C_F c_{0}^{(1,\mbox{\tiny F})}\,,\qquad
  c_{\lambda_1}^{(1)} = C_F c_{\lambda_1}^{(1,\mbox{\tiny F})}\,,\nonumber\\[2mm]
  c_{\lambda_2}^{(1)} &= C_F\left(c_{\lambda_2}^{(1,\mbox{\tiny F})} +
    \Delta c_{\lambda_2}^{(1,\mbox{\tiny F})}\right)+C_A\left(c_{\lambda_2}^{(1,\mbox{\tiny A})} +
    \Delta c_{\lambda_2}^{(1,\mbox{\tiny A})}\right)\,,
\end{align}
our final results are given by
\begin{align}
  c_{0}^{(1,\mbox{\tiny F})} &= -\left(5+\frac{4\pi^2}{3}\right)\delta(1-z) -
  7\left[\frac{1}{1-z}\right]_+\nonumber\\[1mm]
  &\hspace{4.5mm}-\!4\left[\frac{\ln(1-z)}{1-z}\right]_++7+z-2z^2-2 (1+z)\ln(1-z) -
  4\,c_0^{(0)}\ln\frac{\mu}{m_b}\,,\\[3mm]
  c_{\lambda_1}^{(1,\mbox{\tiny F})} &= -\frac{2}{3}\left(15+2\pi ^2\right)\delta(1-z) -
  \frac{4}{3}\left(3-\pi ^2\right)\delta^\prime(1-z)\nonumber\\[1mm]
  &\hspace{4.5mm}-\!\left(\frac{7}{6}-\frac{4\pi^2}{9}\right)\delta^{\prime\prime}(1-z) +
  \frac{2}{3}\left[\frac{1}{(1-z)^3}\right]_+-\frac{5}{3}\left[\frac{1}{(1-z)^2}\right]_+\nonumber\\[1mm]
  &\hspace{4.5mm}-\!9\left[\frac{1}{1-z}\right]_++\frac{8}{3}\left[\frac{\ln(1-z)}{(1-z)^3}\right]_+ -
  4\left[\frac{\ln(1-z)}{(1-z)^2}\right]_+\nonumber\\[1mm]
  &\hspace{4.5mm}-\!4\left[\frac{\ln(1-z)}{1-z}\right]_++10+\frac{7}{3}\,z-\frac{2}{3}\,(4+3z)\ln(1-z) -
  4\,c_{\lambda_1}^{(0)}\ln\frac{\mu}{m_b}\,,\\[3mm]
  c_{\lambda_2}^{(1,\mbox{\tiny F})} &= \left(41+\frac{20\pi^2}{3}\right)\delta(1-z) +
  4\left(1+\pi ^2\right)\delta^\prime(1-z)\nonumber\\[1mm]
  &\hspace{4.5mm}-\!9\left[\frac{1}{(1-z)^2}\right]_++15\left[\frac{1}{1-z}\right]_+ -
  12\left[\frac{\ln(1-z)}{(1-z)^2}\right]_+\nonumber\\[1mm]
  &\hspace{4.5mm}+\!20\left[\frac{\ln(1-z)}{1-z}\right]_+-6+55 z+2(22-17 z)\ln(1-z) -
  4\,c_{\lambda_2}^{(0)}\ln\frac{\mu}{m_b}\,,\\[3mm]
  c_{\lambda_2}^{(1,\mbox{\tiny A})} &= 2\left(7-\frac{8\pi^2}{3}\right)\delta(1-z)+4\,\delta^\prime(1-z) -
  4\left[\frac{1}{(1-z)^2}\right]_+\nonumber\\[1mm]
  &\hspace{4.5mm}+\!2\left[\frac{1}{1-z}\right]_+ -
  16\left[\frac{\ln(1-z)}{1-z}\right]_++2-6z+4 (1+3 z)\ln(1-z)\,.
\end{align}
The contributions $\Delta c_{\lambda_2}^{(1,\mbox{\tiny F})}$ and $\Delta c_{\lambda_2}^{(1,\mbox{\tiny A})}$
are a consequence of the application of the equation of motion of the effective theory in the evaluation of the
matrix elements of the operators given in (\ref{eq::operators}).
Their explicit expressions read
\begin{equation}
  \Delta c_{\lambda_2}^{(1,\mbox{\tiny F})} = 2\,c_{\lambda_2}^{(0)}\,,\qquad
  \Delta c_{\lambda_2}^{(1,\mbox{\tiny A})} = 2\left(1+\ln\frac{\mu}{m_b}\right)c_{\lambda_2}^{(0)}.
\end{equation}
We note that the coefficient $c_0^{(1)}$ agrees with the well-known result of \cite{Ali:1990tj}. The
coefficients $c_{\lambda_1}^{(1)}$ and $c_{\lambda_2}^{(1)}$ have been calculated here for the first time. For the
first one we confirm the expected relation
\begin{equation}
  \int_0^1\!dz\,c_0^{(n)} = -\int_0^1\!dz\,c_{\lambda_1}^{(n)},\qquad n=0,1,2,\dots\,.
\end{equation}
Not surprisingly, $c^{(1)}_{\lambda_{1,2}}$ 
diverge more strongly than $c^{(1)}_0$ at the endpoint. The 
plus-distributions  introduced above follow the prescription 
\begin{align}
  &\int_0^1\!dz\left[\frac{\ln^n(1-z)}{(1-z)^m}\right]_+f(z)\nonumber\\[1mm]
  &\qquad =\int_0^1\!dz\,\frac{\ln^n(1-z)}{(1-z)^m}\left\{f(z) -
  \sum_{p=0}^{m-1}(-1)^p\frac{(1-z)^p}{p!}\left[\frac{\partial^pf(z)}{\partial z^p}\right]_{z=1}\right\}\,,
\end{align}
where $f(z)$ is an arbitrary test function which is regular at $z=1$, and $n\ge 0$, $m\ge 1$. In case the
integration does not include the endpoint, we have ($c<1$)
\begin{equation}
  \int_0^c\!dz\left[\frac{\ln^n(1-z)}{(1-z)^m}\right]_+f(z) =
  \int_0^c\!dz\,\frac{\ln^n(1-z)}{(1-z)^m} f(z)\,.
\end{equation}
We remark that the $\mu$-dependence of $c_{0}^{(1,\mbox{\tiny F})}$ has its origin in the $\MS$ renormalization of
the electromagnetic dipole operator. The same is true for the $\mu$-dependence of
$c_{\lambda_1}^{(1,\mbox{\tiny F})}$ and
$c_{\lambda_2}^{(1,\mbox{\tiny F})}$. It reflects the fact that $\lambda_1$ is not renormalized to all orders in
perturbation theory. On the other hand, for the coefficient of $\lambda_2$ there is an additional $\mu$-dependence
present in $c_{\lambda_2}^{(1,\mbox{\tiny A})}$. It originates in the $\MS$ renormalization of $\lambda_2$. Indeed,
using the renormalization group equation
\begin{equation}
  \mu\frac{d}{d\mu}\,c_{\lambda_2}^{(0)}(\mu) =
  \frac{\alpha_s(\mu)}{4\pi}\,\gamma^{(0)}_{\lambda_2}c_{\lambda_2}^{(0)}(\mu)\,,
\end{equation}
where $\gamma^{(0)}_{\lambda_2}=2C_A$ \cite{Falk:1990pz}, we recover the $\mu$-dependence
of $\Delta c_{\lambda_2}^{(1,\mbox{\tiny A})}$.

In the remainder of this section we discuss the numerical impact of the new contributions on the total decay
rate of $\bar B\to X_s\gamma$ as well as on its first and second moments. For the 
$O_7$ contribution to the  total decay rate  we find (using numerical values for $C_F$
and $C_A$)
\begin{align}\label{rate}
  \Gamma_{77}|_{E_\gamma>E_0} &= \int^1_{z0}\!dz\,\frac{d\Gamma_{77}}{dz} =
  \Gamma_{77}^{(0)}\bigg[1+\frac{\lambda_1-9\lambda_2(\mu)}{2m^2_b} +
    \frac{\alpha_s(\mu)}{4\pi}\bigg(\frac{16}{9}\bigg[4-\pi^2-3\ln\frac{\mu}{m_b}\bigg]\nonumber\\[2mm]
  &\hspace{-8mm}-\!\frac{8}{3}\ln^2(1-z_0)-\frac{4}{3}\left(10-2z_0-z_0^2\right)\ln(1-z_0) -
    \frac{4}{9}z_0\left(30+3z_0-2 z_0^2\right)\nonumber\\[2mm]
  &\hspace{-8mm}+\!\frac{\lambda_1}{2 m^2_b}\bigg\{
    \frac{16}{9}\bigg[4-\pi^2-3\ln\frac{\mu}{m_b}\bigg]-\frac{8}{3}\ln^2(1-z_0)-
    \frac{4z_0\left(30-63z_0+31z_0^2+5z_0^3\right)}{9(1-z_0)^2}\nonumber\\[2mm]
  &\hspace{-8mm}-\!\frac{4\left(30-72z_0+51z_0^2-2z_0^3-3z_0^4\right)}{9(1-z_0)^2}
    \,\ln(1-z_0)\bigg\}+\frac{\lambda_2(\mu)}{2m^2_b}\bigg\{-\frac{2}{9}\bigg[87+32\pi^2\nonumber\\[2mm]
  &\hspace{-8mm}+\!27\ln\frac{\mu}{m_b}\bigg]-\frac{32}{3}\ln^2(1-z_0) +
    \frac{2\left(162-244z_0+113z_0^2-7z_0^3\right)}{3(1-z_0)}\,\ln(1-z_0)\nonumber\\[2mm]
  &\hspace{-8mm}+\!\frac{2z_0\left(54-49z_0+15z_0^2\right)}{1-z_0}\bigg\}\bigg)
    \bigg] + \dots\,.
\end{align}
The ellipses denote higher order terms in $\alpha_s$ and $\Lambda/m_b$,  and we used
$z_0=2E_0/m_b$.  
In order to get a rough estimate of the size of the power-corrections at
$O(\alpha_s)$ we set $\mu=m_b$ and use
the numerical values $\alpha_s(m_b)=0.22$, $m_b=4.6\,{\rm GeV}$,
$\lambda_1=-0.4\,{\rm GeV}^2$ and
$\lambda_2=0.12\,{\rm GeV}^2$ to obtain
$\Gamma_{77}|_{E_\gamma>1.8\,{\rm GeV}}/\Gamma_{77}^{(0)}=0.763-0.007=0.756$\,, a $-0.9$\% effect.
In the intermediate step we have singled out the new $O(\alpha_s\Lambda^2/m_b^2)$ contributions.
The effect of the new corrections on the rate varies with the cut, from $-0.4\%$ at $E_0=0$ to
$-0.9\%$ at $E_0=1.8$\,GeV. For values of $E_0>1.8$\,GeV the corrections to the rate are significant,
a $-3$\% effect for $E_0=2$\,GeV, however, such high values of $E_0$ are well outside
the range of applicability of the local OPE. Moreover, at high $E_0$ part of the new effect
is implicitly contained in the approach of \cite{benson}.

\noindent The truncated $n$-th moment  is defined through
\begin{equation}
  \langle E_\gamma^n\rangle_{E_\gamma>E_0} =
  \left(\frac{m_b}{2}\right)^n\left.\left(\int_{z_0}^1\!dz\,z^n\,\frac{d\Gamma_{77}}{dz}\right)\right/
  \left(\int_{z_0}^1\!dz\,\frac{d\Gamma_{77}}{dz}\right)\,.
\end{equation}
After expanding in $\alpha_s$, the first moment reads
\begin{align}\label{mom1}
  \langle E_\gamma\rangle_{E_\gamma>E_0} &= \frac{m_b}{2}\bigg[1-\frac{\lambda_1+3\lambda_2(\mu)}{2m^2_b}
  +\frac{\alpha_s(\mu)}{4\pi}\bigg(\!-\frac{46}{27}+\frac{8}{9}\left(8-9z_0+z_0^3\right)\ln(1-z_0)\nonumber\\[2mm]
  &\hspace{-11mm}+\!\frac{2}{27}\,z_0\left(96-60z_0-22z_0^2+9z_0^3\right)+\frac{\lambda_1}{2 m^2_b}\bigg\{
    \!-\frac{2z_0\left(96-120z_0-52z_0^2+25 z_0^3-9 z_0^4\right)}{27(1-z_0)}\nonumber\\[2mm]
  &\hspace{-11mm}+\!\frac{46}{27}-\frac{16\left(4-7 z_0+z_0^3\right)}{9(1-z_0)}\,\ln(1-z_0)\bigg\}
    +\frac{\lambda_2(\mu)}{2m^2_b}\bigg\{\!-\frac{1130}{27}-18\ln\frac{\mu}{m_b} +
    \frac{4}{9}\left(80-33z_0\right.\nonumber\\[2mm]
  &\hspace{-11mm}-\!\left.81 z_0^2+25z_0^3\right)\ln(1-z_0) +
    \frac{2}{27}\,z_0\left(480+42z_0-425z_0^2+81z_0^3\right)\bigg\}\bigg)\bigg]+\dots\,,
\end{align}
while the second central moment is given by
\begin{align}\label{mom2}
  \langle E_\gamma^2\rangle-\langle E_\gamma\rangle^2\big|_{E_\gamma>E_0} &= -\frac{\lambda_1}{12}
    +\frac{\alpha_s(\mu)}{4\pi}\bigg\{\frac{m_b^2}{4}\left[-\frac{2}{9}\,(1-z_0)^2\left(17-2z_0-3z_0^2\right)
    \ln(1-z_0)\right.\nonumber\\[2mm]
  &\hspace{-28mm}+\!\left.\frac{1}{270}\,(1-z_0)^2\left(61-898z_0-207z_0^2+144z_0^3\right)\right] +
    \frac{\lambda_1}{8}\bigg[\frac{8}{27}\left(2-27z_0+7z_0^3\right)\ln(1-z_0)\nonumber\\[2mm]
  &\hspace{-28mm}+\!\frac{4}{405}\left(169+60z_0-780z_0^2-385z_0^3+225z_0^4-54z_0^5\right)\bigg] +
    \frac{\lambda_2(\mu)}{8}\bigg[\frac{1}{9}\,\left(43-84z_0\right.\nonumber\\[2mm]
  &\hspace{-28mm}+\!\left.258z_0^2-292z_0^3+75z_0^4\right)\ln(1-z_0) -
      \frac{1}{540}\left(707-2580z_0+3750z_0^2-13820z_0^3\right.\nonumber\\[2mm]
  &\hspace{-28mm}+\!\left.14535z_0^4-2592 z_0^5\right)\bigg]\bigg\} + \dots\,.
\end{align}
Using the same numerical input as above we obtain, for $E_0=1.8\,{\rm GeV}$,
$\langle E_\gamma\rangle = (m_b/2)(0.978-0.001)=2.246\, {\rm GeV}$ and
$\langle E_\gamma^2\rangle-\langle E_\gamma\rangle^2 = 0.0422- 0.0036 = 0.0386\,{\rm GeV}^2$,
where we have again singled out the contribution of $O(\alpha_s\Lambda^2/m_b^2)$ in the
intermediate steps. Their effect on the truncated first and second central moment is of around $-0.1\%$
and $-8.5\%$,
respectively. For $E_0\in[0,1.8]$\,GeV, the effect on the first moment varies between $-0.2$\% and $-0.1$\%,
whereas that on the second central moment varies between $-3.5\%$ and $-8.5\%$.

\begin{figure}[t]
  \begin{center}
  \begin{tabular}{c@{\hspace{0.8cm}}c}  
    \epsfig{figure=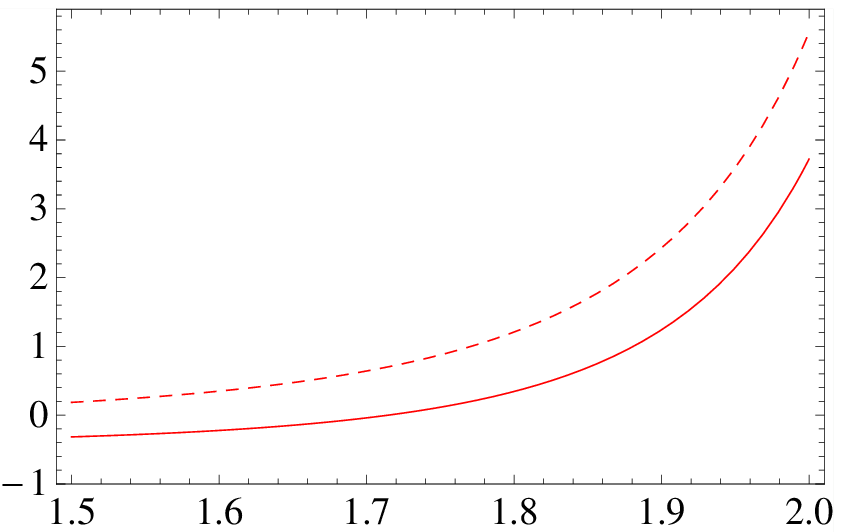,height=4.7cm} &
    \epsfig{figure=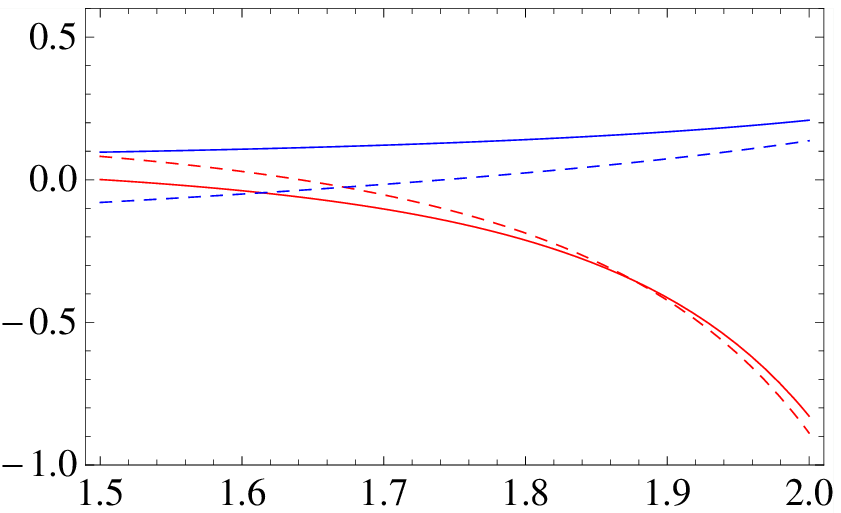,height=4.7cm}\\[-4.5cm]
    \hspace*{-5cm}{\small $\times 0.01$} &
    \hspace*{-4.6cm}{\small $\times 0.01$}\\[3.8cm]
    \hspace*{.5cm}{\small $E_0$} &
    \hspace*{.6cm}{\small $E_0$}
  \end{tabular}
  \caption{\sl NLO coefficients of $\lambda_{1}$ (solid curves) and its leading approximation in $\Delta/m_b$
    (dashed curves) for the decay rate (left panel), the first moment (right panel, lower red curves) and second
    moment (right panel, upper blue curves) as a function of $E_0$.
}\label{plot:expansion}
 \end{center}
\end{figure}

It is interesting to compare the $O(\as)$ coefficients of $\lambda_1$ in (\ref{rate}), (\ref{mom1}) and
(\ref{mom2}) with the results of \cite{neubert}, where the corrections of $O(\alpha_s\lambda_1/\Delta^2)$ with $\Delta=m_b(1-z_0)$ have been computed for the cut rate and the first two moments.
 Expanding our results in $\Delta/m_b=1-z_0$ and keeping only
the leading term we reproduce the results of \cite{neubert}.
Fig.~\ref{plot:expansion} summarizes the numerical relevance of
the terms suppressed by powers of $\Delta/m_b$ in the $O(\as)$ coefficients of $\lambda_1/(2 m_b^2)$ for the same numerical input  used above. 
In the cut rate and in the second moment,
the leading approximation  deviates  by roughly +50\%
and $-35$\%, respectively, already at $E_0=2$GeV. 
In the first moment, the leading approximation is within roughly 10\%
of the complete result down to E = 1.8 GeV.
In conclusion, the range
%The leading approximation in the decay rate and in the second central moment
%deviates for $E_0=2$\,GeV already by roughly $+50$\% and $-35$\%, respectively, from the complete result. For
%the first moment the situation is much better. Here, the leading approximation stays within $+12$\% of the complete result down to $E_0=1.8$\,GeV.
%These results are not surprising, as the purely perturbative corrections follow a similar pattern at
%$O(\alpha_s)$ and $O(\alpha_s^2)$ (check and cite Melnikov-Mitov and our paper).
%We conclude that the range 
of applicability of the leading order approximation in the expansion in
$\Delta/m_b$ is clearly restricted to the region $E_0>2$\,GeV, where Sudakov logarithms become dominant, see for instance \cite{Melnikov}.

\begin{figure}[t]
  \begin{center}
  \vspace*{-1mm}
  \begin{tabular}{c@{\hspace{0.8cm}}c}
    \epsfig{figure=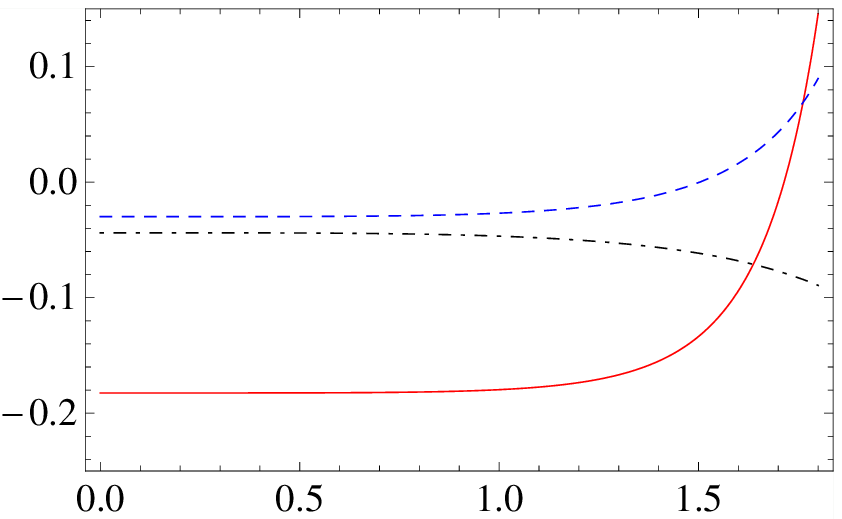,height=4.8cm} &
    \epsfig{figure=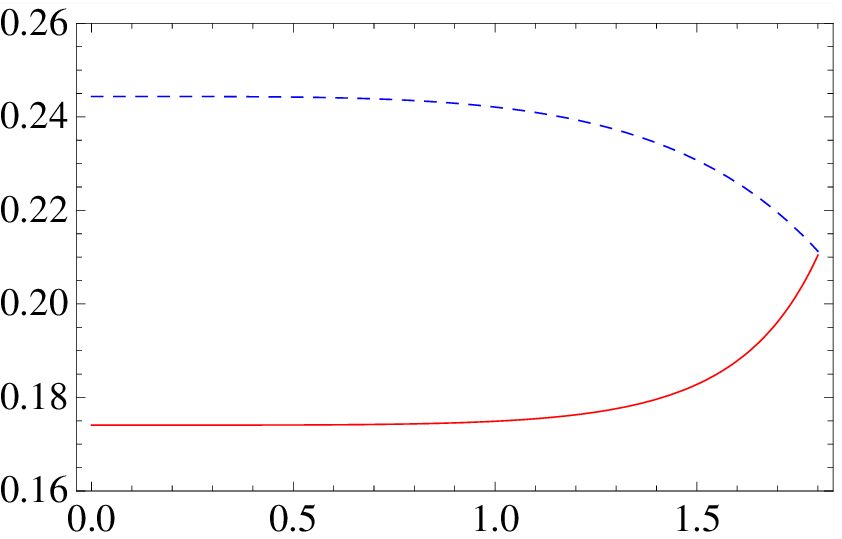,height=4.8cm}\\[-2mm]
    \hspace*{.5cm}{\small $E_0$} &
    \hspace*{.6cm}{\small $E_0$}
  \end{tabular}
  \caption{\sl Ratio of NLO to leading order coefficients of $\lambda_{1}$ (left) and $\lambda_2$ (right) in the
    rate (red solid curves), the first moment (blue dashed curves) and the second moment (black dash-dotted curve)
    as a function of $E_0$.
}\label{plot:coefficients}
 \end{center}
\end{figure}

Fig.~\ref{plot:coefficients} shows the ratios of NLO to leading order coefficients of $\lambda_{1,2}$ in the rate
and in the first two moments as a function of the cut $E_0$ using the same input as above. The NLO corrections to
$\lambda_{2}$ are close to 20\%. Note that in the right panel we have not shown a curve for the second central
moment since $\lambda_{2}$ has a vanishing leading order coefficient.

The second moment represents a powerful constraint on the kinetic 
expectation value $\mu_\pi^2= -\lambda_1+O(1/m_b)$. 
Since the $O(\alpha_s)$ correction decreases its coefficient by 5 to 9\% in 
the range of cuts between 0 and 1.8\,GeV, while the $O(\alpha_s\lambda_2/m_b^2)$ 
corrections are much smaller, we expect to extract a higher value of $\mu_\pi^2$ from radiative moments once the
new corrections are included. Indeed, using $\as$ at a more appropriate scale of order 1-2\,{\rm GeV},
the extracted $\mu_\pi^2$ gets shifted by approximately $+10$\%.
Analogously, the small correction
to the first moment leads to a roughly 10\,MeV positive shift in $m_b$.

All the above expressions refer to the on-shell scheme for $m_b$ and $\lambda_1$, which is inherent in the HQET
calculation. In practical applications however one  adopts short-distance definitions of these 
parameters, as in the kinetic scheme \cite{kinetic}. In this scheme the new corrections are identical, but they
additionally induce small $O(\alpha_s^2\mu_{\rm kin}^2/m_b^2)$
contributions. Therefore, the above rough estimates hold in the kinetic scheme as well. A complete
phenomenological analysis of the moments, including all the available contributions, will be presented elsewhere.

\section{Conclusions}

We have computed the NLO contributions to the Wilson coefficients of 
dimension five operators relevant for inclusive radiative B decays. Our 
results allow for a more precise evaluation of the moments of the photon 
distribution and will improve the  determination of $m_b$ and of the 
kinetic expectation value, $\mu_\pi^2$, from radiative moments.
We have estimated that the new contributions shift the value of $\mu_\pi^2$
extracted from the radiative moments by approximately $+10$\% and that of 
$m_b$ by roughly $+10$\,MeV. The effect on the $\bar B\to X_s\gamma$ rate is below 1\% for $E_0<1.8$\,GeV.

We have performed the calculation analytically, using an off-shell 
matching procedure, a  method that can  be applied to 
inclusive semileptonic decays as well. The $O(\alpha_s\mu_\pi^2/m_b^2)$  corrections to
the moments of $B\to X_c \ell \nu$ have 
been computed numerically \cite{Becher:2007tk}, however, the $O(\alpha_s\mu_G^2/m_b^2)$ corrections are
not yet known. We also believe that an analytical result might be easier to implement in the fitting codes.

\subsection*{\normalsize Acknowledgements}
P.G.~is grateful to G.~Ridolfi for collaboration at the early stage of this work and to N.~Uraltsev for many
useful discussions. We also thank M.~Misiak for relevant communications. 
Work supported  by the EU's Marie-Curie Research Training Network under contract
MRTN-CT-2006-035505 `Tools and Precision Calculations for Physics Discoveries at Colliders'.

\end{document}